\documentclass[final,1p,times,twocolumn]{elsarticle}

\usepackage{amssymb}
\usepackage{amsmath}

\usepackage{float}
\usepackage{multirow}
\usepackage{booktabs}
\usepackage{tabularx}
\usepackage{lineno}
\usepackage{color}
\usepackage{url}

\newcommand{\cm}{cm$^{-1}$}

\newcommand{\CO}[2]{$^{1#1}$C$^{1#2}$O}
\newcommand{\COO}[2]{$^{1#1}$C$^{1#2}$O$_{2}$}
\newcommand{\OCO}[3]{$^{1#1}$O$^{1#2}$C$^{1#3}$O}
\newcommand{\Marvel}{\textsc{Marvel}}
\newcommand{\TROVE}{\textsc{TROVE }}

\journal{Journal of Molecular Spectroscopy}

\begin{document}

\begin{frontmatter}



\title{Machine learning isotope shifts in molecular energy levels}


\author[1]{Marco G. Barnfield}
\author[1,2]{Oleg L. Polyansky}
\author[1]{Sergei N. Yurchenko}
\author[1]{Jonathan Tennyson}

\affiliation[1]{organization={Department of Physics and Astronomy, University College London}, city={London}, postcode={WC1E 6BT}, country={UK}}
\affiliation[2]{organization={Institute of Applied Physics, Russian Academy of Sciences}, addressline={Ulyanov Street 46}, city={Nizhny Novgorod}, postcode={603950}, country={Russia}}

\begin{abstract}
Recent advances in the use of High-Resolution Cross-Correlation Spectroscopy (HRCCS) to detect molecular species  in exoplanet atmospheres, presents a new challenge for the accuracy of reference spectroscopic line lists. While parent isotopologues of key atmospheric tracers are often well-characterized, minor isotopologues, crucial for diagnosing planetary formation histories and evolution, suffer from a scarcity of experimental data, often leading to reliance on less accurate theoretical predictions.
In this work, a comprehensive machine learning framework is designed to mitigate these inaccuracies by modelling the residual errors of the isotopologue extrapolation (IE) method used within the ExoMol project. A fully connected neural network architecture for carbon dioxide (CO$_2$) is shown to predict energy corrections with high fidelity, reducing the mean absolute error (MAE) relative to the original IE approach for more than 87\% of the levels when benchmarked against empirical (\Marvel) energies. Furthermore, development of a novel hybrid, molecule-aware transfer learning architecture is presented that successfully propagates correction patterns from the data-rich CO$_2$ system to the data-poor carbon monoxide (CO) system. This transfer learning approach yields MAE improvements in over 93\% of CO samples, demonstrating that physical correction factors related to isotopic substitution can be generalized across chemically related molecular systems. Updated and improved line lists are presented for 11 CO$_2$ isotopologues and energy levels for excited states of CO isotopologues are predicted.
The methodology establishes a scalable, data-driven paradigm for refining molecular line lists, helping to bridge the gap between theoretical calculations and experimental precision.
\end{abstract}

\begin{highlights}
\item Machine learning is used to improve energy level predictions for minor isotopologues.
\item For CO$_2$ more than 91\% of the isotopologue levels are improved by this procedure.
\item FOr CO more than 93\% of the isotopologue levels are improved by this procedure.
\item Improved line list for CO$_2$ and energy levels for CO are presented.
\end{highlights}


\end{frontmatter}

\section{Introduction}
\label{Introduction}
The field of exoplanetary science has undergone a rapid and transformative evolution over the past three decades. Since the detection of 51~Pegasi~b in 1995 \cite{95MaQu.exo}, which marked the first confirmation of a planet orbiting a main-sequence star other than our Sun, the catalogue of confirmed exoplanets has expanded rapidly. The number of confirmed exoplanets has now passed 6000, a milestone achieved through the tireless operations of space-based missions such as Kepler \cite{10BoKoBa.exogen} and the Transiting Exoplanet Survey Satellite (TESS) \cite{TESS.exo}, alongside ground-based radial velocity surveys \cite{10RiBaMc.PH3}. This initial era of discovery, focused primarily on demographics and orbital dynamics, has now given way to a new phase: atmospheric characterization.

Beyond addressing long-standing questions about habitability and the potential for life, the chemical composition of an exoplanetary atmosphere also provides a fossil record of the planet’s formation. It encodes information about where the planet formed within the protoplanetary disk, how it migrated to its current orbit, and the nature of its interaction with the host star \cite{19Madhus.exo}. For instance, the carbon-to-oxygen (C/O) ratio is widely used as a diagnostic of whether a planet formed beyond the snow lines of key volatile species such as water, carbon dioxide, or carbon monoxide \cite{22MoMoBi.CO}.  For this two complementary spectroscopic techniques are actively being pursued \cite{jt940}.

Transit Spectroscopy has been the primary tool for this chemical forensic work. By analysing the wavelength-dependent absorption or emission of light from the planet, astronomers can identify the unique spectral fingerprints of atmospheric molecules and retrieve temperature–pressure profiles, chemical abundances, and dynamical information such as winds and rotation rates \cite{15BaAiIr.exo}. The recent launch and successful deployment of the James Webb Space Telescope (JWST) has provided the community with high signal-to-noise, moderate- to high-resolution infrared spectra of exoplanet atmospheres from space, enabling multi-molecule detections and joint retrievals of composition and thermal structure \cite{24WiBeLi.CO}. 

In parallel, the next generation of $\geq$ 30~m ground-based facilities, such as the European Southern Observatory’s Extremely Large Telescope (ELT) \cite{15SnKoBi.PH3}, is under construction and will host high-resolution spectrographs operating at the resolving powers $R=\lambda / \Delta \lambda \approx 10^{5}$. These instruments will exploit high-resolution cross-correlation spectroscopy (HRCCS), wherein by leveraging the large orbital Doppler shifts of the planet relative to the quasi-stationary stellar and telluric spectra, HRCCS effectively separates the planetary component, thus delivering the highest effective spectral resolving power among current techniques for exoplanet atmosphere characterization \cite{25Snellen}. This enables constraints on molecular abundances, atmospheric dynamics, and even 3D structure that are inaccessible at low resolution.

In the regime of high-resolution spectroscopy, the requirements for reference data precision become exceptionally stringent. While standard, computed line lists are often sufficient for lower-resolution transit studies, where broad absorption bands dominate the signal, HRCCS relies on matching thousands of individual spectral lines. At $R\geq10^{5}$, the small shifts in energy levels caused by isotopic substitution become resolvable and, consequently, scientifically critical. Isotopologues are molecules that differ only in their isotopic composition, such as the substitution of a carbon-12 atom for carbon-13 in carbon monoxide (\CO{2}{6} vs. \CO{3}{6}). Despite typically low abundances, these species provide invaluable diagnostics. The ratio of $^{12}$C to $^{13}$C, for example, is not affected by chemical processing in the same way as elemental ratios and can serve as a robust tracer of the primary isotope reservoir from which the planet accreted. Recent observational successes, such as the detection of \CO{3}{6} in the atmosphere of the super-Jupiter TYC~8998-760-1~b \cite{21YaSnBo.CO} and the hot-Jupiter WASP-39~b \cite{23EsLoAd.CO} using cross-correlation methods, confirm that isotopologue detections are not merely theoretical in exoplanet atmospheres, but practically achievable with current technology.

However, the detectability of isotopologues is severely limited by the quality of the underlying spectroscopic data. Simulation studies have demonstrated that HRCCS is acutely sensitive to inaccuracies in line positions \cite{19BrLixx.exo}. If the theoretical template used for cross-correlation is shifted even slightly from the true physical spectrum, the cross-correlation signal is degraded or lost entirely. Biases in the line lists can lead to erroneous retrievals of atmospheric abundances, temperatures, and velocities. Consequently, the promise of isotopic diagnostics in exoplanets depends directly on our ability to model the spectroscopy of minor isotopologues with the same fidelity as the parent species.

To address the spectroscopic needs of astrophysics, the ExoMol project  was established with the mandate to produce comprehensive line lists for all molecules likely to be observable in hot atmospheres \cite{jt528}. The ExoMol methodology combines sophisticated \textit{ab initio} quantum mechanical calculations with laboratory data to generate line lists that are complete (often covering billions of transitions) and accurate. 

A critical component in refining these theoretical models is the use of empirical energy levels derived from a large set of experimental data. The Measured Active Rotational-Vibrational Energy Levels (\Marvel) algorithm is a powerful and widely used method for this purpose \cite{jt412}. \Marvel\ systematically processes all available assigned transitions for a given molecule, often collated from dozens of different experimental studies.
These transitions are used to construct a spectroscopic network, where the quantum states are represented as nodes (or vertices) and the observed transitions between them are the links (or edges). This network-based approach is used to validate the input transitions, check for inconsistencies, and identify outliers or misassigned lines from the experimental data. Following this validation, a weighted least-squares fit is performed on the network to determine a self-consistent set of empirical energy levels with robustly determined uncertainties. These \textsc{Marvel}-derived energy levels provide the high-quality benchmark data used to refine the \textit{ab initio} potential energy surface in the ExoMol procedure, ensuring that the final line lists meet the stringent accuracy demands of modern astrophysics \cite{jt939}.

Despite the success of ExoMol, which currently provides comprehensive line lists for almost hundred  parent  molecules, a significant data gap exists for minor isotopologues. Experimental data for these species is often sparse, with laboratory spectroscopy historically focused on the most abundant isotopologues due to signal-to-noise limitations. Without a dense network of experimental energy levels to refine the theoretical potential energy surfaces (PES), the calculations for minor isotopologues rely on extrapolations that inevitably introduce errors.

The ``isotopologue extrapolation'' (IE) method was introduced by Polyansky \textit{et al}. \cite{jt665} for water isotopologues, then formalized by McKemmish \textit{et al}. \cite{jt948} and applied to TiO, MgO and VO, leading to significant improvements in the accuracy as shown by its use to detect of TiO isotopologues in stellar spectra \cite{jt799}. The method leverages experimental residuals ($E_{\text{exp}} - E_{\text{calc}}$) of a parent isotopologue to correct calculated energy levels of minor isotopologues:
\begin{equation}
    E^{\text{\rm iso}}_{\text{\rm IE}} = E^{\text{\rm iso}}_{\text{\rm Ca}} + \left(E^{\text{\rm parent}}_{\text{\rm Ma}} - E^{\text{\rm parent}}_{\text{\rm Ca}}\right),
    \label{Eq:ie_original}
\end{equation}
where $E^{\text{\rm parent}}_{\text{\rm Ca}}$ is the  variationally calculated energy of the parent isotopologue, $E^{\text{\rm parent}}_{\text{\rm Ma}}$ is the empirically derived \Marvel\ \cite{jt412} energy, $E^{\text{\rm iso}}_{\text{\rm Ca}}$ is the calculated variational energy of the given isotopologue, and $E^{\text{\rm iso}}_{\text{\rm IE}}$ is the extrapolated isotopologue energy. In essence, the residuals in energy between the empirical and calculated line lists are assumed to be constant for all isotopologues \cite{jt665,jt948}. While this is a reasonable first-order approximation, it fails to account for the subtle, mass-dependent effects of the breakdown of the Born-Oppenheimer approximation. These residual errors, while small, are large enough to hamper high-resolution studies.

The limitations of the constant-shift assumption in standard IE become apparent when considering the physical origins of the discrepancy between variational calculations and experimental data. These discrepancies are not random numerical artifacts; rather, they arise primarily from the breakdown of the Born-Oppenheimer (BO) approximation \cite{Schwenke2001}. The BO approximation relies on the separation of electronic and nuclear motion based on their disparate masses. However, neglecting the coupling between these motions introduces specific error terms, adiabatic and non-adiabatic corrections, that scale inversely with nuclear mass and evolve with internal energy. Consequently, the residual encapsulates these neglected physical interactions. Because these effects follow deterministic physical laws dependent on quantum numbers and mass, the residuals are structured and predictable. This physical determinism is the crucial prerequisite that renders these residuals ``learnable", thus providing the theoretical justification for a machine learning approach.

Machine Learning (ML), specifically deep learning, has emerged as a powerful paradigm for modelling complex, non-linear physical phenomena \cite{15HaBiRa.ML,25WeMa.ML}. The confluence of large-scale spectroscopic datasets and advanced computational techniques presents a unique opportunity. The ExoMol project provides an extensive database of transitions and energy levels that can be utilised by ML, with uses already in line broadening by Guest \textit{et al.} \cite{jt919} and automated quantum number assignment in the CO$_2$ ``Dozen'' line list \cite{jt999}.

In this paper a shift from purely physical extrapolation to a data-driven correction framework is proposed from the original IE method. By training neural networks to map quantum numbers and physical descriptors to these residuals, high-accuracy energy corrections can be predicted for unmeasured states in the minor isotopologues of carbon dioxide (CO$_2$) and carbon monoxide (CO), two molecules of immense astrophysical importance. Furthermore, transfer learning is explored by using the knowledge learned from the data-rich CO$_2$ system to improve predictions for the data-poor CO system. This work aims to establish a robust, scalable, and transferable pipeline for generating high-fidelity spectroscopic data, thereby supporting the next generation of exoplanet observations.

\section{Method}
Although in principle machine learning (ML) can be employed to directly predict the energy levels of minor isotopologues from those of the parent isotopologue, in practice this approach yields larger residuals than the original IE method. Instead, this work presents a proof of concept that residuals of the original IE method against empirical (\textsc{Marvel}) levels can be used in a ML framework. We show that the residual at each energy level ($\Delta E_i$) is learned to produce corrections to energy levels across all isotopologues as follows 
\begin{equation}
    E^{\mathrm{\rm iso}}_{\mathrm{\rm ML}} = E^{\mathrm{\rm iso}}_{\mathrm{\rm IE}} + \Delta E^{\mathrm{\rm iso}}_i.
    \label{Eq:ie_correction}
\end{equation}
The residual itself is defined by
\begin{equation}
    \Delta E^{\mathrm{\rm iso}}_i = E^{\mathrm{\rm iso}}_{\mathrm{\rm Ma}} - E^{\mathrm{\rm iso}}_{\mathrm{\rm IE}}.
    \label{Eq:residual_def}
\end{equation}

\subsection{Data Curation \& Preparation}
\label{sec:data}
The CO$_2$ data was derived from the ``Dozen'' line list \cite{jt999}, a comprehensive update to the ExoMol database for carbon dioxide. The isotopologues' energy levels were extracted in a series of \textsc{Marvel} studies which considered the parent species \COO{2}{6} as well all the 11 stable minor isotopologues \cite{jt925,jt932,jt955,jt963,jt973,jt974,jt983,jt984}, such as \COO{3}{6}, \OCO{6}{2}{8}, and \COO{3}{8} which we denote 636, 628, 638, respectively, below. For the parent \CO{2}{6} and the five stable CO isotopologues (\CO{2}{7}, \CO{2}{8}, \CO{3}{6}, \CO{3}{7}, \CO{3}{8}, henceforth denoted 27, 28, 36, 27 and 38), updated energy levels derived from recent \textsc{Marvel} analyses were adopted \cite{jt961,jt1008}. For both CO$_2$ and CO, the \textsc{Marvel} datasets contain a significant number of validated, empirical energy levels: 8268 levels spanning energies up to 20654 \cm\ with $J\leq 118$ for \COO{2}{6} and 2293 levels spanning energies up to 67~148 \cm\ with $v\leq 41$ and $J\leq 123$ for \CO{2}{6}; specific quantum number coverage and energy limits for the individual minor isotopologues are detailed in Tables~\ref{tab:data_CO2} and \ref{tab:data_CO}. The quantity of available \textsc{Marvel}-derived energy levels for the minor CO isotopologues is significantly lower than for CO$_2$ (Tables~\ref{tab:data_CO2} and \ref{tab:data_CO}); this data scarcity necessitates the transfer learning approach described in Section~\ref{transfer_learning}. Only energy levels possessing values for all four necessary components ($E^{\rm parent}_{\rm Ma}$, $E^{\rm parent}_{\rm Ca}$, $E^{\rm iso}_{\rm Ca}$, $E^{\rm iso}_{\rm Ma}$) were retained in our ML procedure. This intersection is necessary to calculate both the input features (the original IE prediction) and the target label (the true residual to the \textsc{Marvel}-derived energy level) for supervised learning.

\begin{table}[ht]
\centering
\caption{Summary of the empirical \textsc{Marvel} data available for the minor isotopologues of CO$_2$, detailing the maximum vibrational ($v_i^{\text{max}}$ in Herzberg notation) and rotational ($J^{\text{max}}$) quantum numbers, maximum energy ($E^{\text{max}}$ in cm$^{-1}$), and total count of levels used.}
\label{tab:data_CO2}
\setlength{\tabcolsep}{5pt}
\begin{tabular}{l r r r r r r}
\hline
\textbf{Isotopologue} & \boldmath{$v_1^{\text{max}}$} & \boldmath{$v_2^{\text{max}}$} & \boldmath{$v_3^{\text{max}}$} & \boldmath{$J^{\text{max}}$} & \textbf{$E^{\text{max}}$ (cm$^{-1}$)} & \textbf{Count} \\ \hline
\OCO{6}{2}{7} (627) & 4 & 5 & 5 & 109 & 13469.81 & 4639 \\
\OCO{6}{2}{8} (628) & 6 & 5 & 5 & 111 & 13465.76 & 6008 \\
\COO{3}{6} (636)    & 5 & 18 & 9 & 113 & 19985.94 & 6918 \\
\OCO{6}{3}{7} (637) & 4 & 4 & 4 & 99  & 9178.13  & 2714 \\
\OCO{6}{3}{8} (638) & 4 & 5 & 5 & 105 & 9758.55  & 3881 \\
\COO{2}{7} (727)    & 4 & 4 & 5 & 89  & 13169.70 & 1851 \\
\OCO{7}{2}{8} (728) & 4 & 3 & 4 & 90  & 9105.74  & 2799 \\
\COO{3}{7} (737)    & 2 & 5 & 3 & 56  & 7175.70  & 541  \\
\OCO{7}{3}{8} (738) & 3 & 2 & 3 & 83  & 8594.01  & 963  \\
\COO{2}{8} (828)    & 6 & 0 & 3 & 100 & 13514.70 & 4663 \\
\COO{3}{8} (838)    & 4 & 6 & 3 & 91  & 9467.74  & 1703 \\ \hline
\end{tabular}
\end{table}

\begin{table}[ht]
\centering
\caption{Summary of the empirical \textsc{Marvel} data available for the minor isotopologues of CO, detailing the maximum vibrational ($v^{\text{max}}$) and rotational ($J^{\text{max}}$) quantum numbers, maximum energy ($E^{\text{max}}$ in cm$^{-1}$), and total count of levels used.}
\label{tab:data_CO}
\setlength{\tabcolsep}{8pt}
\begin{tabular}{l r r r r}
\hline
\textbf{Isotopologue} & \boldmath{$v^{\text{max}}$} & \boldmath{$J^{\text{max}}$} & \textbf{$E^{\text{max}}$ (cm$^{-1}$)} & \textbf{Count} \\ \hline
\CO{2}{7} (27) & 4 & 21 & 8689.70 & 33 \\
\CO{2}{8} (28) & 22 & 38 & 40467.91 & 498 \\
\CO{3}{6} (36) & 27 & 45 & 48166.08 & 737 \\
\CO{3}{7} (37) & 1 & 22 & 2964.58 & 45 \\
\CO{3}{8} (38) & 13 & 30 & 25080.73 & 345 \\ \hline
\end{tabular}
\end{table}

\subsection{Feature Engineering \& Importance}
\label{features}
To enable the neural network to learn the physics of the residuals, a rich ML feature set was constructed representing the quantum state and isotopic properties of each energy level. The features (Table~\ref{tab:features_CO2}) were divided into continuous variables (standardized) and categorical variables (encoded). Notably, states with $J+v_3$ odd, denoted B' and B''   symmetries by TROVE, appear in the 727 and 737 isotopologues; these levels are excluded from the training set as there are no corresponding experimental levels within the parent isotopologue 626 due to nuclear spin statistics.

\begin{table}[H]
\caption{Feature set used for the CO$_2$ Neural Network.}
\label{tab:features_CO2}
\centering
\begin{tabularx}{\textwidth}{@{} l X @{}} 
\toprule
\textbf{Feature Name} & \textbf{Description} \\ 
\midrule

\multicolumn{2}{l}{\textit{Energies}} \\ 
\addlinespace[2pt]
$E^{\rm iso}_{\rm Ca}$    & Calculated energy (minor isotopologue) \\
$E^{\rm parent}_{\rm Ca}$ & Calculated energy (parent isotopologue) \\
$E^{\rm parent}_{\rm Ma}$ & Calculated energy (parent isotopologue) \\
$E^{\rm iso}_{\rm IE}$    & Original IE calculated energy (minor isotopologue) \\ 
\addlinespace[10pt]

\multicolumn{2}{l}{\textit{Quantum Numbers}} \\ 
\addlinespace[2pt]
$J$                     & Total rotational quantum number \\
$g_{\rm tot}$           & Total state degeneracy \\
$v_1, v_2, l_2, v_3$    & Normal-mode vibrational quantum numbers (Herzberg notation) \\
$m_1, m_2, l_2, m_3, r$ & AFGL vibrational quantum numbers \\
$t_1, t_2, t_3$         & TROVE local-mode quantum numbers \\ 
\addlinespace[10pt]

\multicolumn{2}{l}{\textit{Reduced Masses}} \\ 
\addlinespace[2pt]
$\mu_1, \mu_2, \mu_3, \mu_{\rm all}$ & Reduced masses (sym., asym., and bend) and respective ratios to the parent isotopologue. See \ref{app:mu_eqns} for equations. \\ 
\addlinespace[10pt]

\multicolumn{2}{l}{\textit{Boolean Flags}} \\ 
\addlinespace[2pt]
Atomic masses & e.g.: Presence of $^{13}$C atom; Presence of $^{16}$O in position 1 \\
e/f           & Parity labels \\
$A'$/$A''$    & TROVE symmetry labels \\ 
\bottomrule
\end{tabularx}
\end{table}

Feature importance was subsequently analysed via test-set permutation scoring to aid interpretability regarding which spectroscopic inputs most strongly influenced the learned CO$_2$ corrections. In this approach, each input feature is randomly shuffled in the held-out data and the resulting degradation in predictive accuracy is measured. Consequently, features whose perturbation most harms performance are identified as the most influential for the model’s corrected CO$_2$ energy predictions. 

\subsection{Network Architecture I: CO$_2$}
\begin{figure}[H]
    \centering
    \includegraphics[width=1\linewidth]{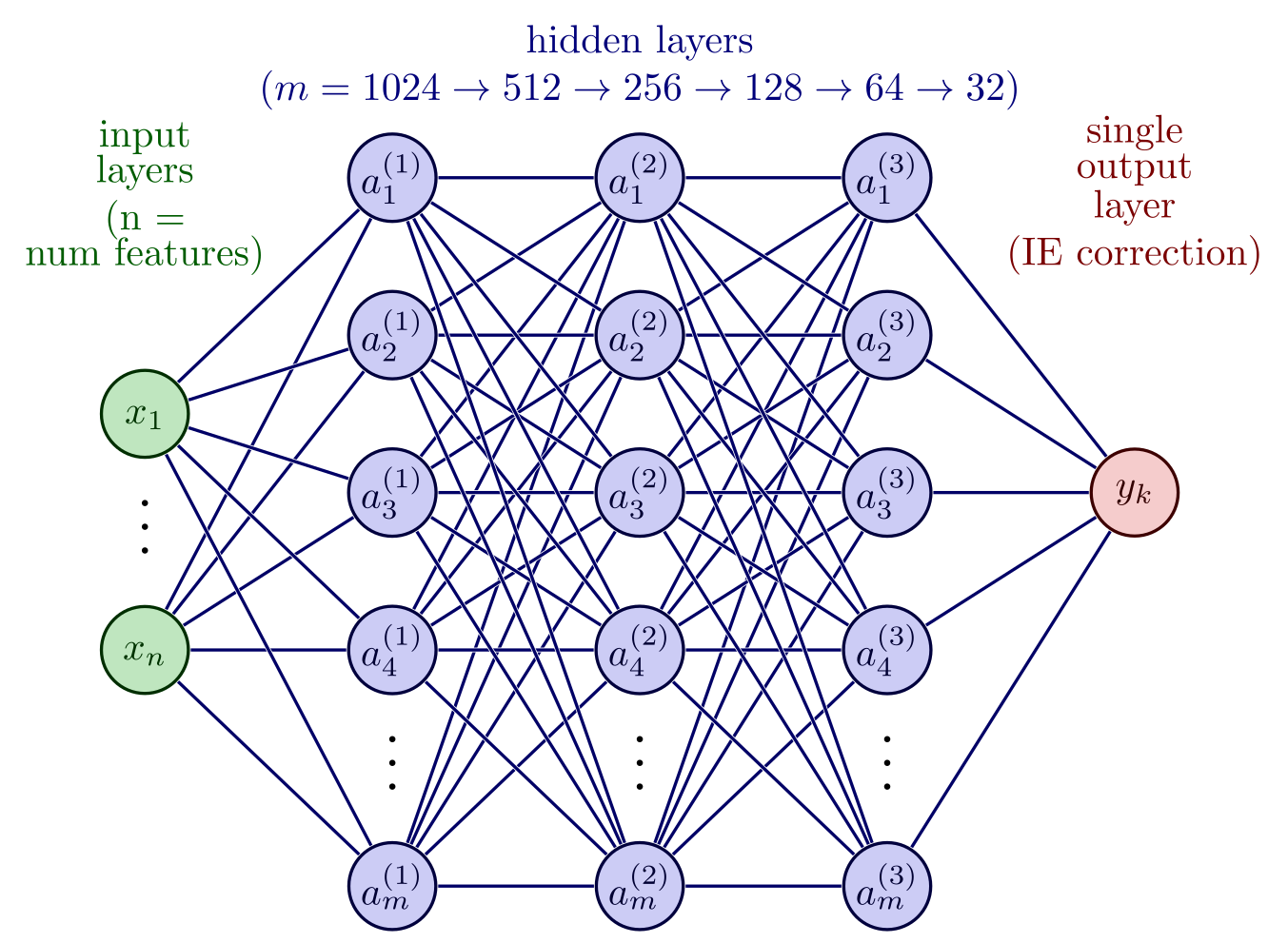}
    \caption{Neural Network structure for CO$_2$ IE corrections.}
    \label{fig:co2_nn_structure}
\end{figure}

For the data-rich CO$_2$ dataset, a feed forward neural network was implemented using the PyTorch \cite{19PaGrMa_Pytorch.ML} library. A full representation of the structure is shown in Fig.~\ref{fig:co2_nn_structure}, containing six hidden layers, each with a decreasing number of units from 1024 to 32, with a single final output layer to predict the correction.

The activation function used was the Gaussian Error Linear Unit (GELU) \cite{23HeGi_GELU.ML}. Unlike the standard Rectified Linear Unit (ReLU) \cite{19Ag_ReLU.ML} which has a sharp discontinuity at zero, GELU, is a smooth, probabilistic activation function. This smoothness is advantageous for regression tasks in physics, where the target function (in this case the energy correction) is continuous and differentiable \cite{23HeGi_GELU.ML}.

Dropout layers were interleaved between the dense layers which randomly zero out a fraction of neurons during training, thus preventing the network from relying too heavily on any single feature or memorizing noise in the training data. This promotes generalization to the unseen energy levels.

The optimizer used was Adam (adaptive moment estimation) \cite{17KiBa_Adam.ML} with a learning rate of $1\times10^{-3}$ and a weight decay of $1\times10^{-6}$. The loss function was a Huber Loss \cite{64Hu_HuberLoss.ML}, which behaves similarly to mean squared error (MSE) for small errors and mean absolute error (MAE) for large errors. Additionally it is robust to outliers, essential given that experimental spectroscopic data can contain misassignments which would skew a pure MSE loss.

The training protocol involved a standard 70/10/20 split between training, validation and test sets, ensuring all isotopologues were represented in each split. Final evaluation used only the untouched test split, reporting loss, root mean square error (RMSE) and MAE. After inference, the model’s predicted correction $\Delta E^{\rm iso}$ was combined with the baseline calculated energy to compute corrected energies (Eq.~\ref{Eq:ie_correction}) and their improvement relative to the uncorrected values ($E_{\rm IE}^{\rm iso}$). 

\subsection{Network Architecture II: Transfer Learning for CO$_2$}
\label{transfer_learning}
\begin{figure}[H]
    \centering
    \includegraphics[width=1\linewidth]{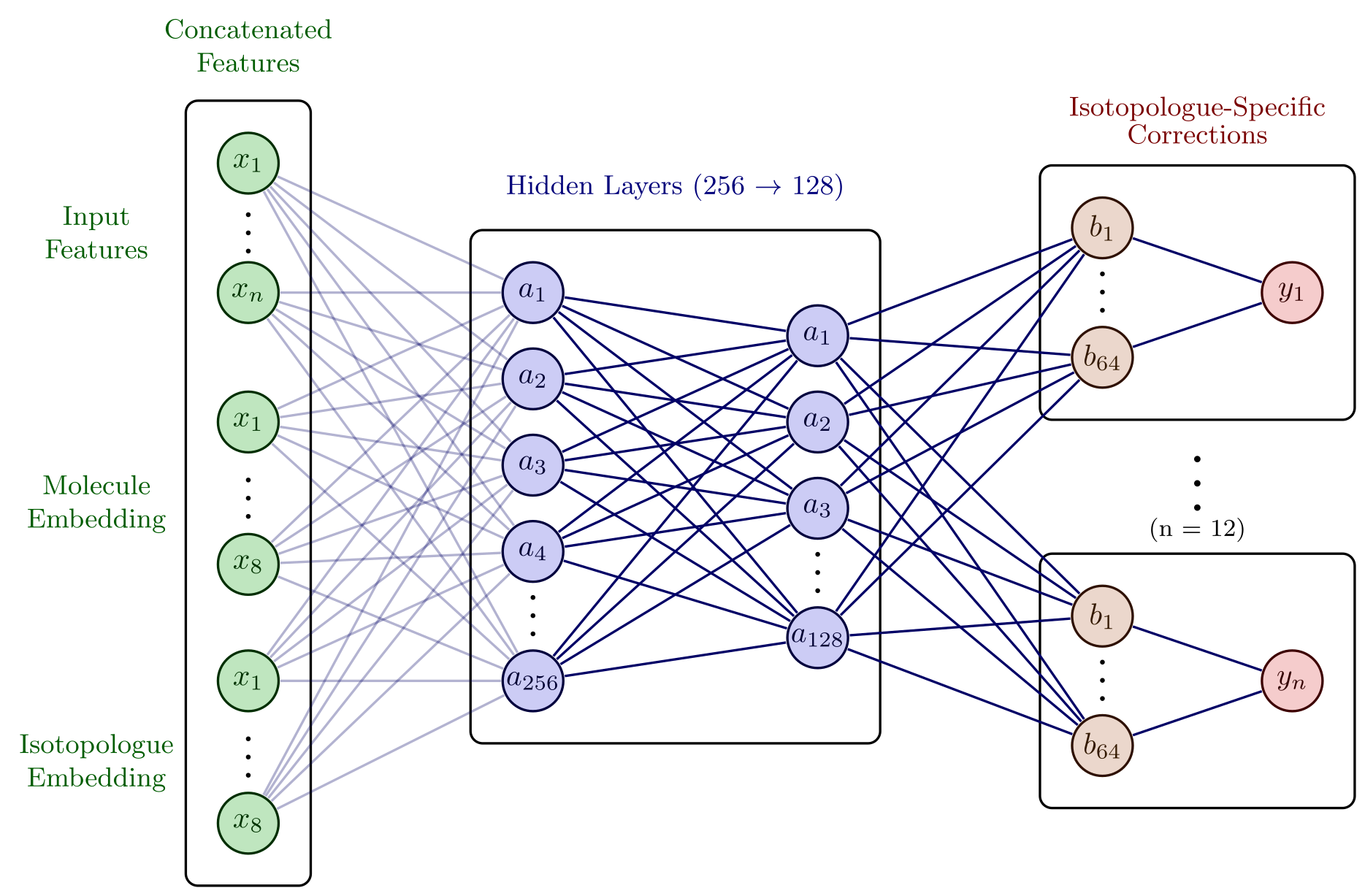}
    \caption{Neural Network structure for CO IE corrections.}
    \label{fig:co_nn_structure}
\end{figure}

A hybrid, molecule-aware architecture is used for CO that allows the network to learn generalized correction patterns for the data-rich CO$_2$ dataset while utilising individual ``adapter heads'' for each isotopologue to capture the specific spectroscopic nuances of CO. This transfer-learning approach ensures that the model benefits from the large scale statistics of CO$_2$ without losing specificity for CO. A full representation of the structure is shown in Fig.~\ref{fig:co_nn_structure}.

The shared trunk now consists of just two dense layers (256 $\to$ 128) with LayerNorm \cite{16BaKiHi_LayerNorm.ML} to normalize the inputs across the features, GELU activations and dropout; a learned gating mechanism blends the shared and isotope-specific outputs to give the final scalar correction $\Delta E^{\rm ML}$ at each isotopologue specific head. This design preserves the ability to transfer broad chemical patterns across data, while still allowing per-isotopologue and per-molecule adjustments when required. Training used Adam optimisation with a lower learning rate of $5\times10^{-4}$ with $1\times 10^{-6}$ weight decay and Huber loss as before. For stability, the training gradients were clipped to a maximum norm of 1.0 and the model output bias was initialised to the mean target of the training subset to speed convergence.

A random weighted sampler was employed to ensure that CO samples were drawn more frequently during training, counteracting CO$_2$ dominating learning and predictions, thus preventing the network from ignoring the minority class. The training strategy also contained adaptive weighting, i.e., if a specific isotopologue's performance lagged, its weight in the loss function was dynamically increased, forcing the network to focus on the ``harder'' cases. Given the small sample size, a stratified 5-fold cross-validation (CV) was also implemented to ensure the performance in all available CO data points was adequate. To further improve reliability and reduce dependence on random initialisation, the complete cross-validation process was repeated across five independent random seeds. For each seed, improvement in prediction accuracy relative to the original IE residuals was quantified using per-isotopologue MAE. The final reported improvement is given as a mean ± standard deviation across seeds, providing a clear measure of both the overall correction achieved and the variability of the results across runs.

\section{Results \& Discussion}
\subsection{Global Performance Summary}
To provide a unified quantitative comparison across both molecular systems, the mean absolute errors (MAEs) for the raw variational calculations from the ``Dozen" line list (Ca) \cite{jt999}, the original isotopologue extrapolation method (IE), and the machine–learning corrected IE energies (ML) are summarised in Table~\ref{tab:global_mae}. The metrics show that, while the original IE method yields only marginal improvement over the raw calculated values, the machine–learning models deliver a substantial reduction in MAE for both CO$_2$ and CO. This demonstrates that the residual–learning approach captures structured, isotopologue–dependent deviations that neither variational calculations nor constant–shift extrapolation adequately model.

\begin{table}[H]
\centering
\caption{Mean absolute errors (MAE) in cm$^{-1}$ for variational calculations (Ca), original isotopologue extrapolation (IE), and ML-corrected-IE (ML) predictions for CO$_2$ and CO energy levels when compared with their respective empirical (\Marvel) levels (previously presented in Tables~\ref{tab:data_CO2} and \ref{tab:data_CO}).
}
\label{tab:global_mae}
\begin{tabular}{lccc}
\hline
\textbf{Molecule} & \textbf{\rm Ca} & \textbf{\rm IE} & \textbf{\rm ML} \\
\hline
CO$_2$ & 0.01395 & 0.01394 & 0.00232 \\
CO      & 0.03007 & 0.02896 & 0.00524 \\
\hline
\end{tabular}
\end{table}

The improvement is especially notable for CO, where the ML correction reduces the MAE by approximately an order of magnitude compared with the original IE method. For CO$_2$, the numerical improvements are smaller because the extensive \Marvel\ dataset already makes the IE predictions highly accurate; however, the ML corrections still deliver a clear and consistent reduction in error beyond what IE alone achieves. This cross–molecule behaviour strengthens the case for employing machine–learning residual correction as a general framework for isotopologue refinement, irrespective of data abundance.

\subsection{CO$_2$}
\begin{figure}[H]
    \centering
    \includegraphics[width=1\linewidth]{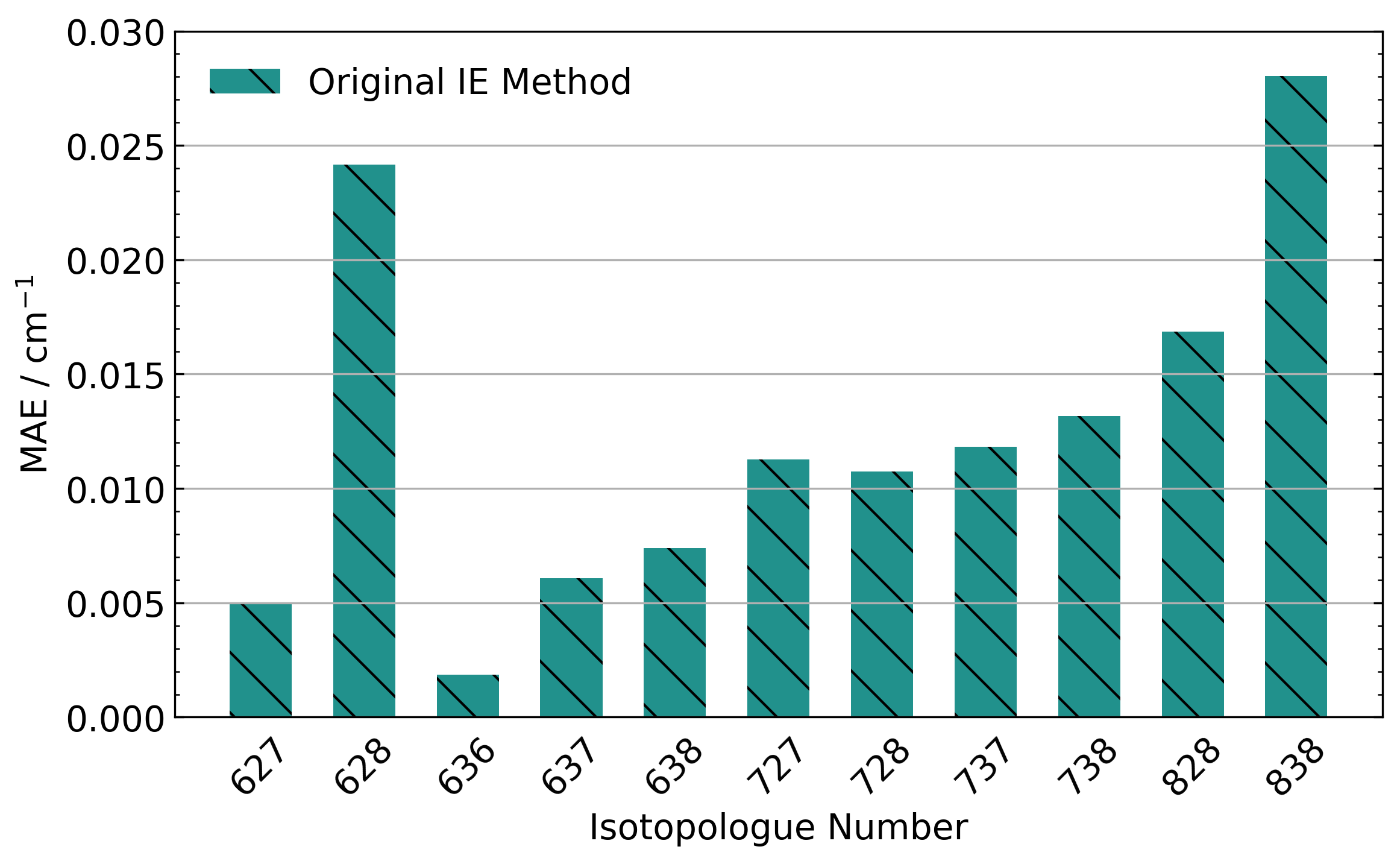}
    \caption{The mean absolute error (MAE) of the original Isotopologue Extrapolation (IE) method for each CO$_2$ minor isotopologue.}
    \label{fig:og_ie_mae_co2}
\end{figure}

While initial training included all eleven minor CO$_2$ isotopologues, the \COO{3}{6} (636) isotopologue was subsequently excluded from the training set. This decision is empirically justified by analzying the baseline performance of the original IE method. As shown in Fig.~\ref{fig:og_ie_mae_co2}, the original IE method yields an MAE for 636 of $<0.002$~cm$^{-1}$, nearly an order of magnitude lower than the problematic species in 628 or 838.

\begin{figure}[H]
    \centering
    \includegraphics[width=1\linewidth]{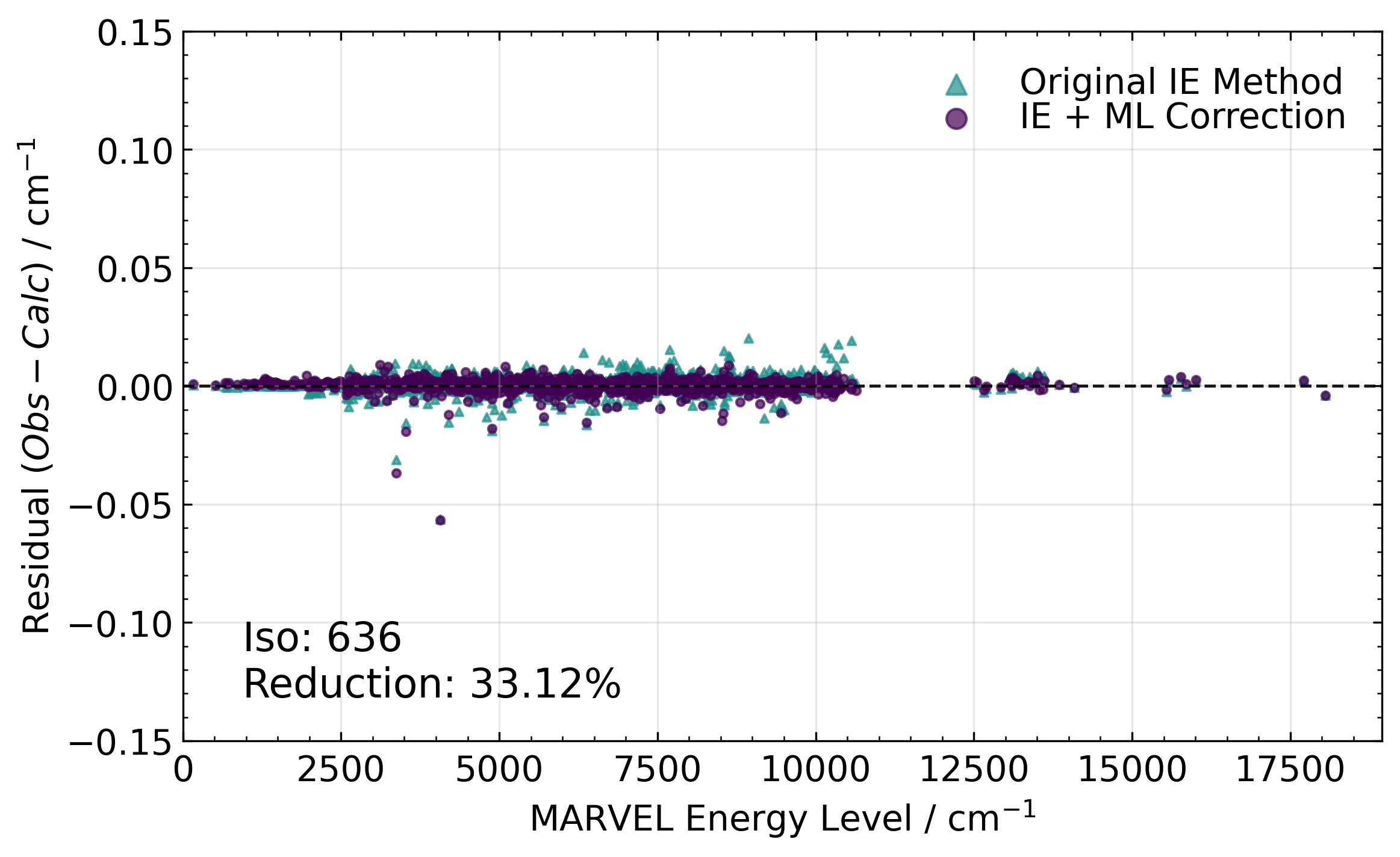}
    \caption{\CO{3}{6} residuals before and after ML correction.}
    \label{fig:636_CO2}
\end{figure}

It is worth noting that when included, the neural network did successfully learn the residual physics for this species, achieving a further 33.12~\% reduction in error as illustrated in the residual distribution in Fig.~\ref{fig:636_CO2}. However, this comparatively marginal gain for an already accurate species came at a significant cost. Because 636 represents a large proportion of the available data ($\approx$18\%), including these low-error targets dominated the loss function. This hampered the network's ability to predict corrections for the remaining isotopologues, resulting in a lower global accuracy. Consequently, 636 was removed to allow the model to focus on the species where the standard approximation fails significantly.

When 636 was included, the overall MAE improvement across isotopologues was 85.93~\%, with 83.14~\% of individual samples showing improvement. After removing 636, the overall MAE improvement increased to 89.27~\%, and the proportion of samples showing improvement rose to 91.62~\% over the original IE method.

\begin{figure}[H]
    \centering
    \includegraphics[width=1\linewidth]{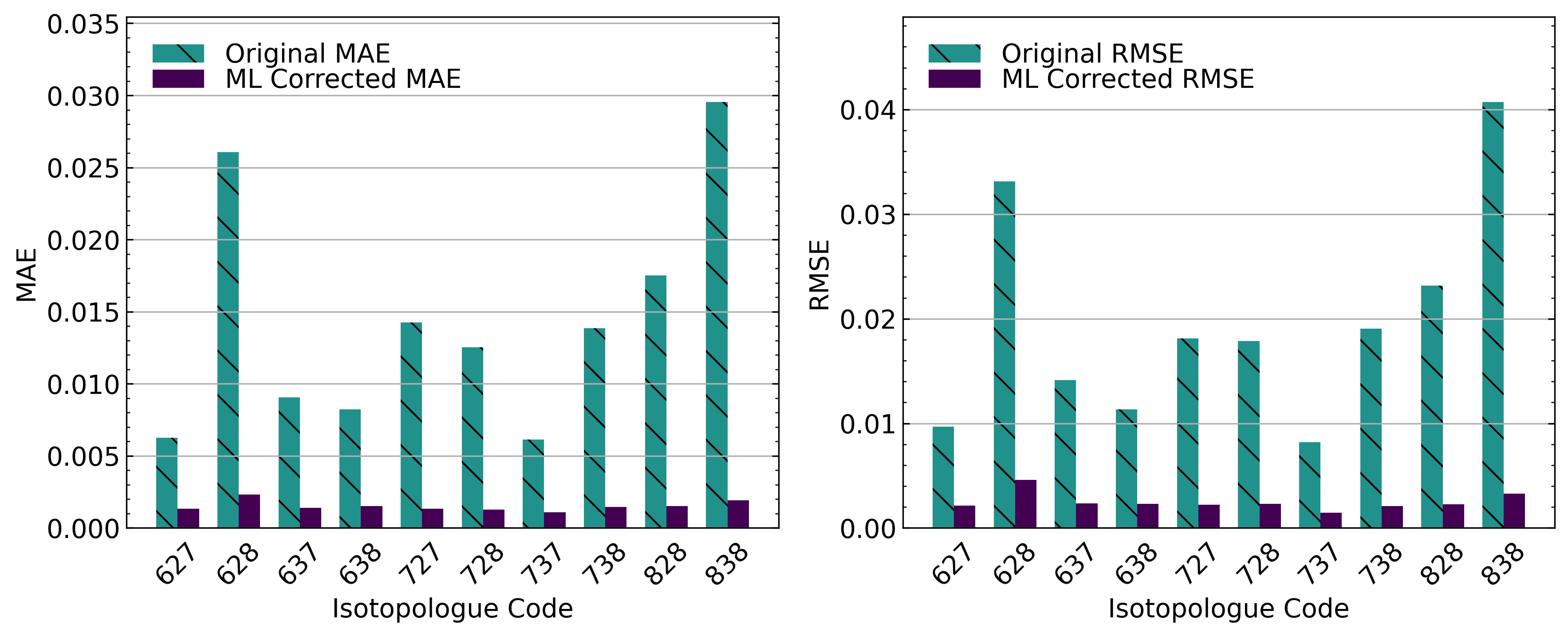}
    \caption{Mean absolute error (MAE) and root mean square error (RMSE) across minor CO$_2$ isotopologues before and after the ML correction.}
    \label{fig:metrics_CO2}
\end{figure}

The overall model performance following this adjustment is shown in Fig.~\ref{fig:metrics_CO2}, which compares the mean absolute error (MAE) and root mean squared error (RMSE) across the remaining CO$_2$ isotopologues before and after neural-network correction. Both metrics decrease substantially following correction, confirming that the model successfully compensates for systematic discrepancies present in the original IE calculations. The reductions are consistent across isotopologues, suggesting that the network captures generalizable correction patterns rather than overfitting to any specific isotopic composition. The greater reduction observed in RMSE compared with MAE indicates that the model is particularly effective at correcting large outliers, which tend to dominate the squared-error measure.

\begin{figure}[H]
    \centering
    \includegraphics[width=1\linewidth]{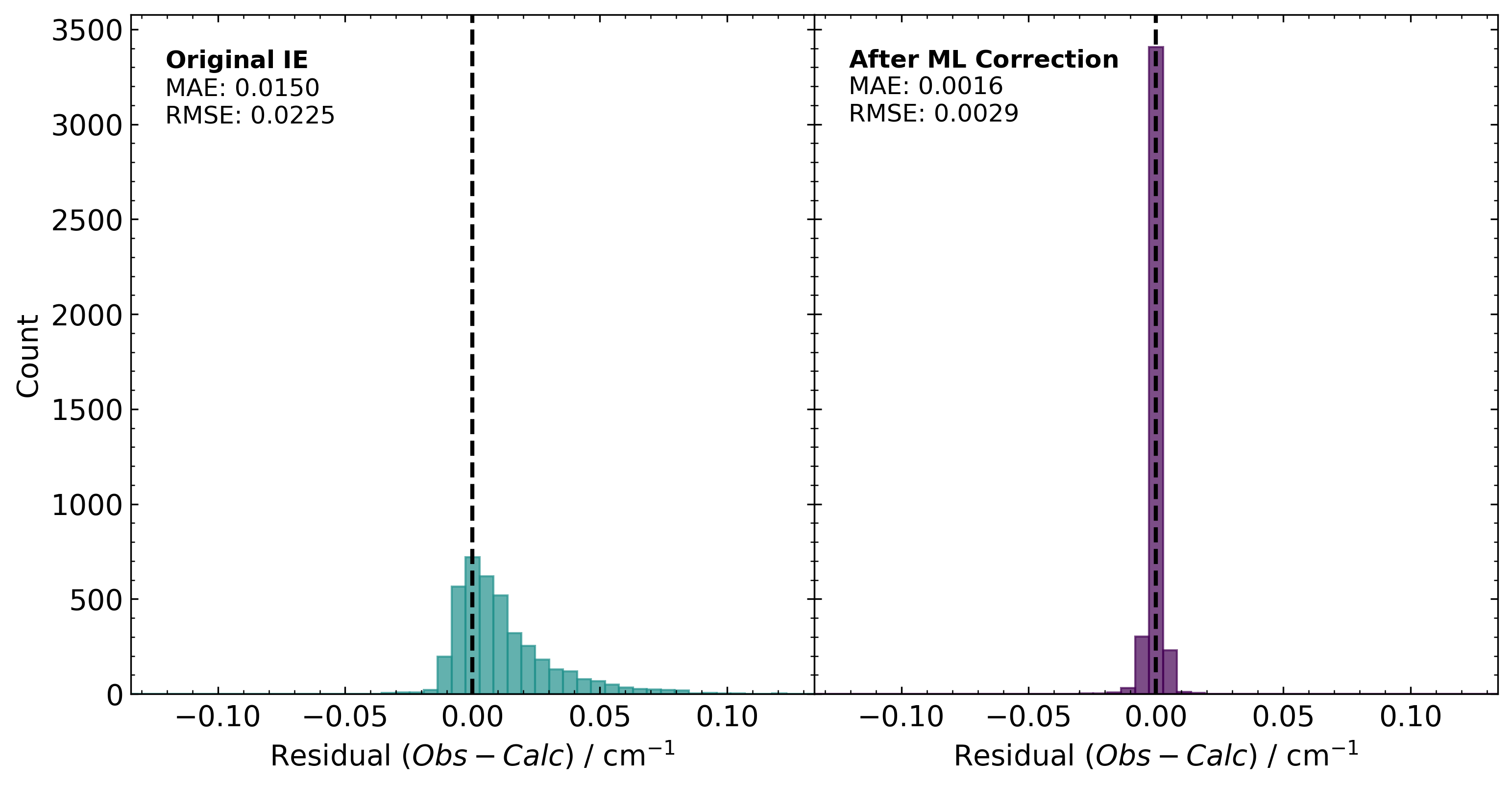}
    \caption{Distribution of residuals for all CO$_2$ isotopologues, representing the discrepancy between empirical \Marvel\ energy levels and the IE-calculated energies before and after ML correction.}
    \label{fig:residuals_CO2}
\end{figure}

\begin{figure}[H]
    \centering
    \includegraphics[width=1\linewidth]{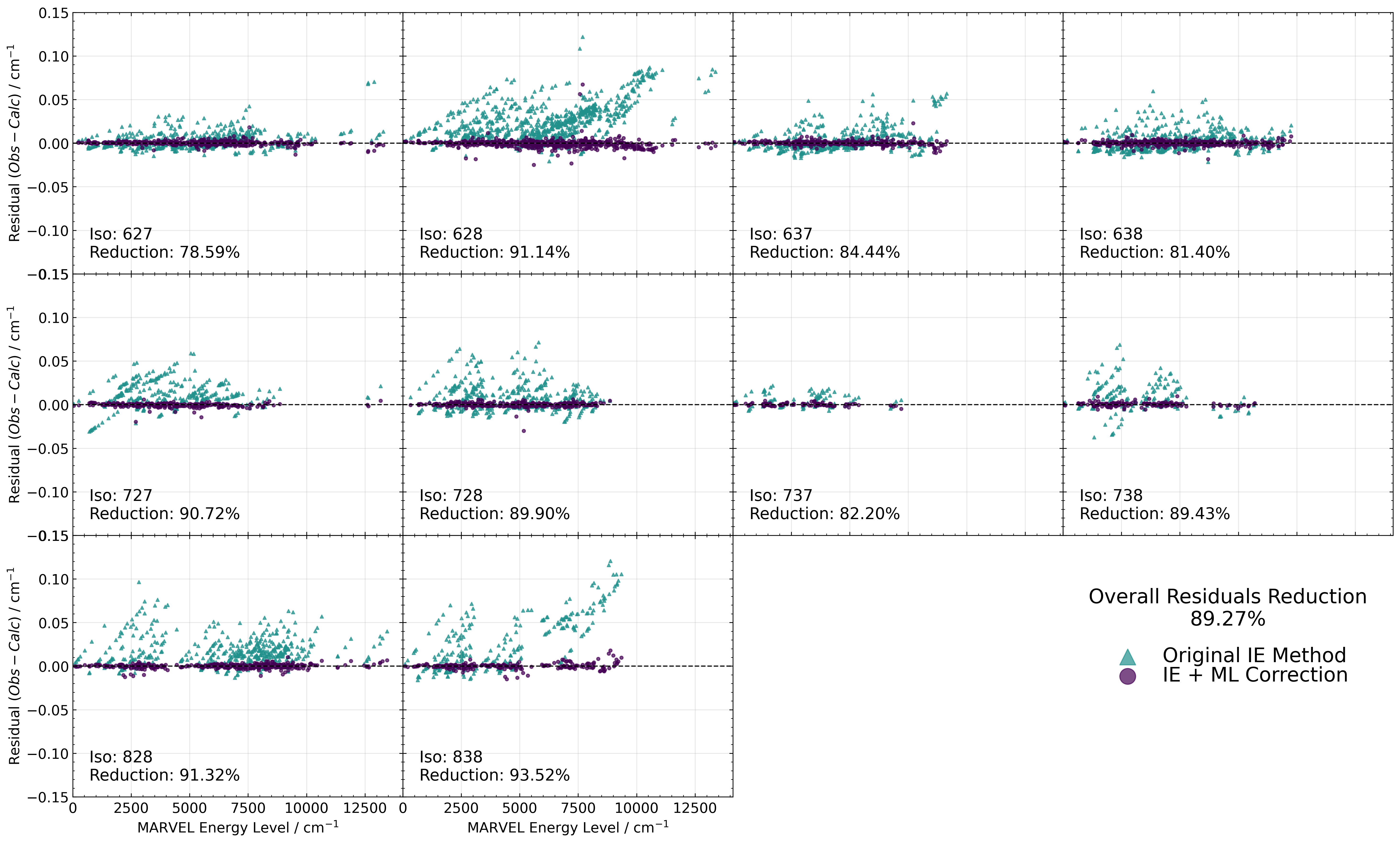}
    \caption{Residuals for individual CO$_2$ isotopologues plotted against \textsc{Marvel} empirical energy levels before and after the ML correction.}
    \label{fig:individual_residuals_CO2}
\end{figure}

The distribution of residuals across all CO$_2$ isotopologues before and after correction in the unseen test set is shown in Fig.~\ref{fig:residuals_CO2}. Prior to correction, residuals are widely spread and biased to the right, indicating that the original IE method systematically overestimates corrections to the calculated values. This behavior can also be observed in the individual isotopologue residual plots shown in Fig.~\ref{fig:individual_residuals_CO2}, with the minor exception of \COO{3}{7} (737), which is routinely underestimated. After correction, the residuals form a much narrower and more symmetric distribution centered around zero, and distinct patterns are no longer discernible within the individual isotopologue subplots. This demonstrates the network's ability to remove systematic bias while substantially reducing errors across the full energy-level range.

\subsubsection{Feature Analysis}
Feature importance was examined using the ablation approach described in Section~\ref{features}, where individual features were removed and the corresponding change in MAE was recorded. Fig.~\ref{tab:feature_imp_CO2} shows that all importance values were positive, indicating that removing any single feature reduced the overall predictive accuracy.
The most substantial contributions to minimizing the mean absolute error (MAE) were derived from the inclusion of the isotopic masses and key spectroscopic predictors, $J$ and vibrational quantum numbers from TROVE, Herzberg, and AFGL. Standard normal-mode vibrational quantum numbers are $v_1, v_2, l_2, v_3$ in Herzberg notation; AFGL ($m_1, m_2, l_2, m_3, r$) refers to the Air Force Geophysics Laboratory notation standard~\cite{81RoYoxx.CO2}; TROVE quantum numbers ($t_1, t_2, t_3$) correspond to local-mode assignments from the variational calculations.

The strong importance of explicit isotopic masses and rovibrational quantum numbers, beyond what is captured by simple reduced-mass scaling, is consistent with the known isotopic dependence of non-adiabatic (Born–Oppenheimer breakdown) corrections in high-precision molecular spectra. Taken together, these results indicate that the network relies on physically meaningful quantities and that the learned corrections encode genuine isotopic dependencies rather than fortuitous correlations.

\begin{table}[H]
\centering
\caption{Feature importance for the CO$_2$ correction network, based on MAE increase upon ablation, highlighting the dominant role of isotopic masses and rovibrational spectroscopic predictors. (Oxygen mass position is denoted in brackets)
}
\label{tab:feature_imp_CO2}
\begin{tabular}{lrllr}
\cline{1-2} \cline{4-5}
Feature                   & Importance &  & Feature               & \multicolumn{1}{l}{Importance} \\ \cline{1-2} \cline{4-5} 
$J$                       & 6.68e-03   &  & $^{12}$C              & 8.02e-04                       \\
$^{17}$O (2)              & 2.84e-03   &  & AFGL l2               & 7.30e-04                       \\
TROVE $v_3$               & 2.26e-03   &  & Symmetry $A_2$        & 6.69e-04                       \\
Herzberg $v_3$            & 2.23e-03   &  & Herzberg l2           & 6.53e-04                       \\
Herzberg $v_1$            & 2.06e-03   &  & $^{18}$O (1)          & 6.06e-04                       \\
$^{13}$C                  & 1.94e-03   &  & Symmetry $A''$        & 5.87e-04                       \\
AFGL $m_3$                & 1.75e-03   &  & AFGL $m_2$            & 5.84e-04                       \\
$E_{\rm Ca}^{\rm parent}$ & 1.70e-03   &  & Symmetry $A'$         & 3.68e-04                       \\
AFGL $m_1$                & 1.66e-03   &  & Symmetry $A_1$        & 3.62e-04                       \\
Herzberg $v_2$            & 1.53e-03   &  & f                     & 1.18e-04                       \\
$E_{\rm Ca}^{\rm iso}$    & 1.43e-03   &  & $\mu_3$               & 7.56e-05                       \\
$E_{\rm Ma}^{\rm parent}$ & 1.33e-03   &  & e                     & 6.39e-05                       \\
$^{17}$O (1)              & 1.28e-03   &  & $\mu_2$               & 2.60e-05                       \\
Trove $v_2$               & 1.20e-03   &  & Trove coefficient     & 1.09e-05                       \\
$^{16}$O (1)              & 1.17e-03   &  & $\mu_1$               & 8.98e-06                       \\
AFGL $r$                  & 1.16e-03   &  & $\mu_3$ ratio         & 4.13e-06                       \\
$\mu_{\rm all}$           & 1.13e-03   &  & $\mu_{\rm all}$ ratio & 2.46e-06                       \\
$g_{\rm tot}$             & 1.12e-03   &  & $\mu_2$ ratio         & 1.69e-06                       \\
$E_{\rm IE}^{\rm iso}$    & 9.80e-04   &  & $\mu_1$ ratio         & 1.01e-06                       \\
Trove $v_1$               & 9.75e-04   &  & $^{16}$O (2)          & 1.00e-06                       \\
$^{18}$O (2)              & 8.08e-04   &  &                       &                                \\ \cline{1-2} \cline{4-5} 
\end{tabular}
\end{table}

\newpage
\subsection{Carbon Monoxide (CO)}
\begin{figure}[H]
    \centering
    \includegraphics[width=1\linewidth]{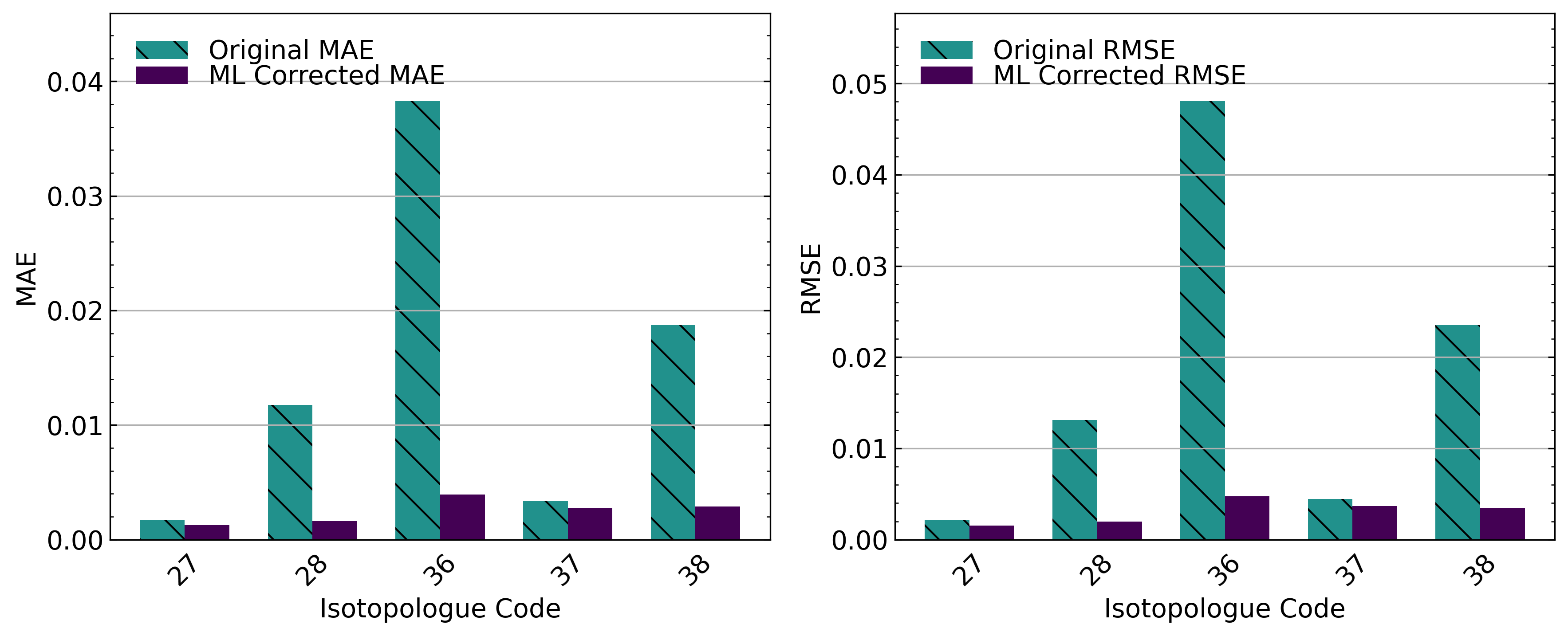}
    \caption{Mean absolute error (MAE) and root mean square error (RMSE) across minor CO isotopologues before and after the ML correction.}
    \label{fig:metrics_CO}
\end{figure}

Much like CO$_2$, the CO network demonstrates a clear improvement in predictive accuracy following neural-network correction, as shown in Fig.~\ref{fig:metrics_CO}, yielding an average MAE improvement of 87.82~\% over the original IE method. In total, 91.37~\% of samples showed improved agreement with experiment after the ML corrections were applied. An example of these improvements is presented in Table~\ref{tab:CO28_improvements}. These results confirm that the hybrid molecule-aware architecture effectively generalises across isotopologues while retaining molecule-specific correction capability.

\begin{table}[ht]
\centering
\caption{Comparison of empirical (\textsc{Marvel}), variational (Calc.), and machine learning corrected (ML) energy levels for the $v=10$ vibrational state of the $^{12}$C$^{18}$O isotopologue (28). All energy and residual values are given in cm$^{-1}$.}
\label{tab:CO28_improvements}
\setlength{\tabcolsep}{5pt}
\begin{tabular}{c c c c c c}
\hline
\textbf{J} & \textbf{\textsc{Marvel} (Ma)} & \textbf{Calc. (Ca)} & \textbf{Resid. (Ma - Ca)} & \textbf{ML} & \textbf{Resid. (Ma - ML)} \\ \hline
0  & 19794.0437 & 19794.0350 & $8.65 \times 10^{-3}$  & 19794.0423 & $1.38 \times 10^{-3}$  \\
1  & 19797.3802 & 19797.3718 & $8.38 \times 10^{-3}$  & 19797.3804 & $-2.01 \times 10^{-4}$ \\
2  & 19804.0539 & 19804.0453 & $8.63 \times 10^{-3}$  & 19804.0562 & $-2.32 \times 10^{-3}$ \\
3  & 19814.0635 & 19814.0551 & $8.42 \times 10^{-3}$  & 19814.0649 & $-1.44 \times 10^{-3}$ \\
4  & 19827.4095 & 19827.4009 & $8.64 \times 10^{-3}$  & 19827.4108 & $-1.29 \times 10^{-3}$ \\
5  & 19844.0905 & 19844.0821 & $8.40 \times 10^{-3}$  & 19844.0913 & $-7.52 \times 10^{-4}$ \\
6  & 19864.1068 & 19864.0981 & $8.64 \times 10^{-3}$  & 19864.1081 & $-1.29 \times 10^{-3}$ \\
7  & 19887.4564 & 19887.4482 & $8.24 \times 10^{-3}$  & 19887.4553 & $1.11 \times 10^{-3}$  \\
8  & 19914.1399 & 19914.1312 & $8.71 \times 10^{-3}$  & 19914.1397 & $2.48 \times 10^{-4}$  \\
9  & 19944.1543 & 19944.1463 & $8.07 \times 10^{-3}$  & 19944.1549 & $-5.64 \times 10^{-4}$ \\
10 & 19977.5008 & 19977.4921 & $8.69 \times 10^{-3}$  & 19977.5013 & $-5.20 \times 10^{-4}$ \\ \hline
\end{tabular}
\end{table}

\begin{figure}[H]
    \centering
    \includegraphics[width=1\linewidth]{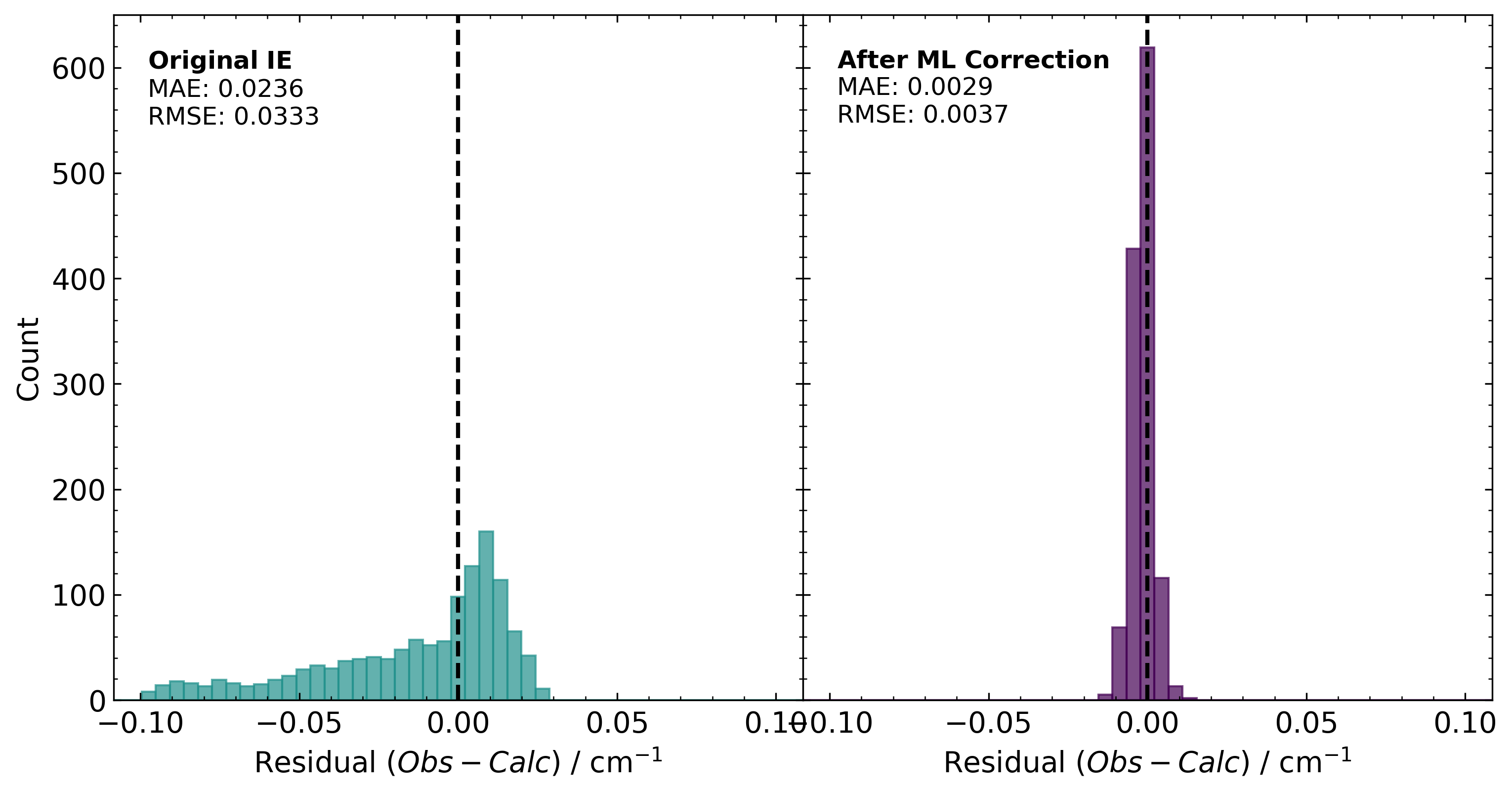}
    \caption{Distribution of residuals for CO isotopologues before and after the ML correction.}
    \label{fig:residuals_CO}
\end{figure}

The residual distributions in Fig.~\ref{fig:residuals_CO} mirror those observed for CO$_2$, narrowing significantly and centring around zero after correction. The network successfully removes the negative bias of the original IE calculations in CO, indicating that the inclusion of CO$_2$ data helped stabilise learning and extend the correction trends to CO. The inset metrics summarise the overall error reduction and confirm the improvement seen across isotopologues.  

\begin{figure}[H]
    \centering
    \includegraphics[width=1\linewidth]{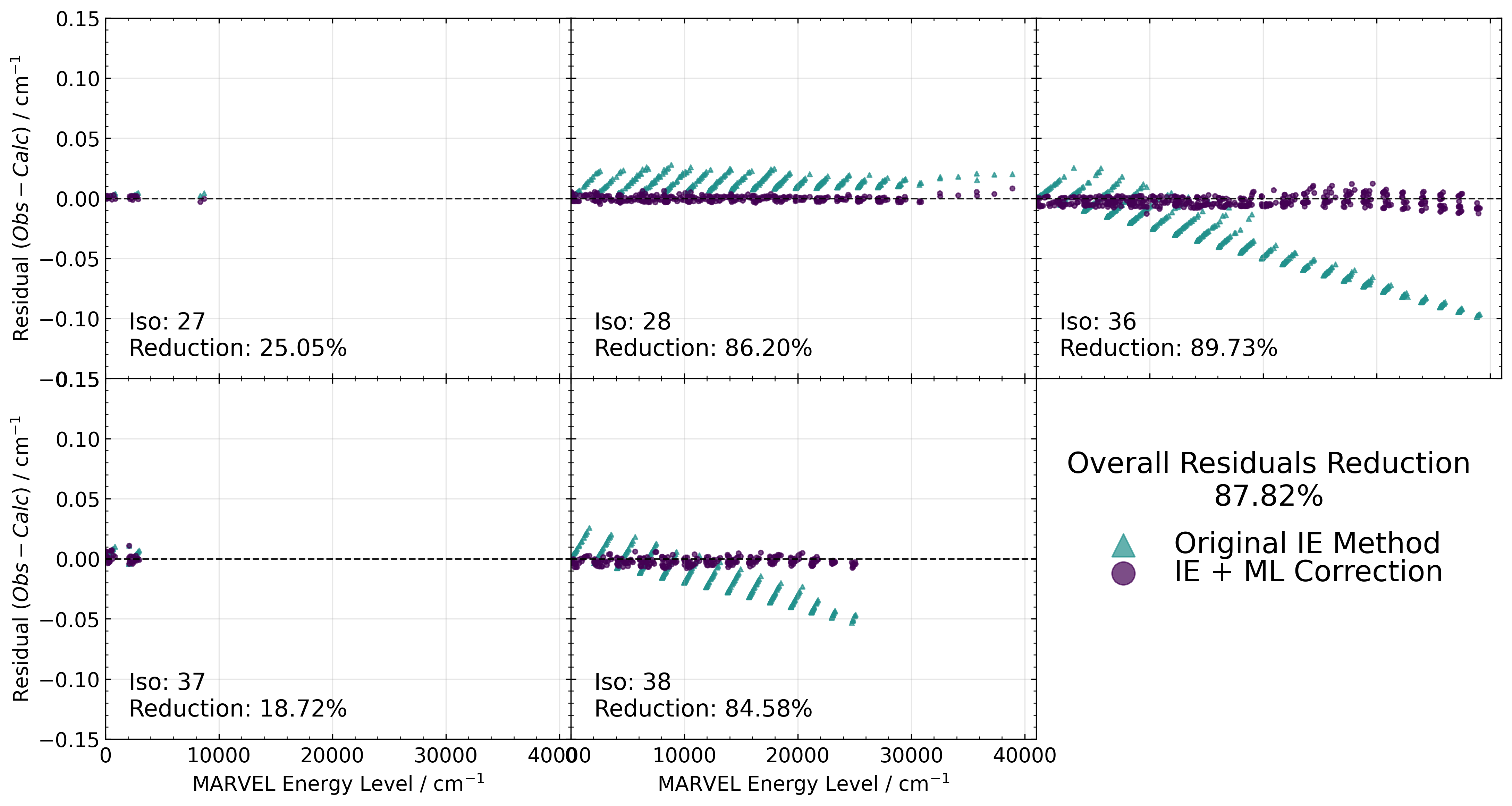}
    \caption{Residuals for individual CO isotopologues plotted against empirical \textsc{Marvel} energy levels before and after the ML correction.}
    \label{fig:individual_residuals_CO}
\end{figure}

Figure~\ref{fig:individual_residuals_CO} presents the residuals for each CO isotopologue plotted against the empirical \textsc{Marvel} energy levels. As with CO$_2$, pre-correction residuals display systematic structure, particularly at higher energies, which largely disappears after correction. The uniform reduction in error across isotopologues demonstrates the success of transferring learned correction behaviour from the CO$_2$ dataset to CO.

\newpage
\subsection{Inference}
Inference for all CO$_2$ isotopologues was performed on calculated energy levels up to 12~500~cm$^{-1}$, consistent with the energy range of the training set. Applying this cutoff resulted in the update of 36~795 energy levels; a breakdown of these totals for each isotopologue is provided in Table \ref{tab:prediction_total_co2}. All levels have been used to update the ``Dozen'' line list \cite{jt999}.

\begin{table}[h]
\centering
\caption{Number of previously variationally calculated energy levels corrected using the IE ML method for each minor CO$_2$ isotopologue.}
\label{tab:prediction_total_co2}
\begin{tabular}{lr}
\hline
\textbf{Isotopologue} & \textbf{No. Energy Levels} \\
\hline
\OCO{6}{2}{7} (627) & 1791 \\
\OCO{6}{2}{8} (628) & 1053 \\
\OCO{6}{3}{7} (637) & 2917 \\
\OCO{6}{3}{8} (638) & 2352 \\
\COO{2}{7} (727) & 7040 \\
\OCO{7}{2}{8} (728) & 2805 \\
\COO{3}{7} (737) & 7041 \\
\OCO{7}{3}{8} (738) & 3764 \\
\COO{2}{8} (828) & 3154  \\
\COO{3}{8} (838) & 4878 \\
\hline
\end{tabular}
\end{table}

Inference for all CO isotopologues was performed up to 40~000~cm$^{-1}$, consistent with the energy range of the training set. Applying this cutoff resulted in the update of 3348 energy levels; a breakdown of these totals for each isotopologue is provided in Table \ref{tab:prediction_total_co}. These CO data will be used for future isotopologue studies within the \textsc{ExoMol} project.

\begin{table}[H]
\centering
\caption{Number of previously variationally calculated energy levels corrected using the IE ML method for each minor CO isotopologue.}
\label{tab:prediction_total_co}
\begin{tabular}{cr}
\hline
\textbf{Isotopologue} & \textbf{No. Energy Levels} \\
\hline
\CO{2}{7} & 1152 \\
\CO{2}{8} & 693   \\
\CO{3}{6} & 545   \\
\CO{3}{7} & 1144 \\
\CO{3}{8} & 844   \\
\hline
\end{tabular}
\end{table}

\section{Conclusion}
The CO$_2$ and CO isotopologue extrapolation correction networks demonstrate that neural networks can effectively learn physically meaningful relationships between isotopic composition and energy-level deviations. Moving beyond traditional methods that rely on simple, global numerical corrections, this work implements a per-energy-level correction scheme. This granular approach resulted in the correction of 36~795 energy levels for CO$_2$ and 3348 energy levels for CO, yielding significant improvements in accuracy over previous uniform scaling methods.

The CO$_2$ model established a robust baseline for isotopologue-specific correction, while the CO network extended this approach by integrating cross-molecular information through a hybrid, molecule-aware architecture. The CO results closely follow the high-accuracy trends of CO$_2$, with the key distinction that the hybrid design enables the effective transfer of learned isotopic correction trends between related molecular systems. This transfer learning approach improves model robustness, accelerates convergence, and enhances correction accuracy even for isotopologues with limited training data.

Together, these results confirm that carefully structured neural networks can both replicate and generalize spectroscopic corrections across families of chemically related molecules. Future work will focus on extending these capabilities to a broader range of molecules, particularly those containing hydrogen atoms where energy level shifts are significantly larger and more heavily influenced by non-Born-Oppenheimer effects. This work ultimately supports the production of more accurate and complete spectroscopic data for use across high-resolution astrophysical applications.

\section*{CRediT authorship contribution statement}
\textbf{Marco G. Barnfield} Writing – Original Draft, Visualization, Validation, Investigation, Formal analysis, Data curation. \textbf{Jonathan Tennyson:} Writing – Review \& Editing, Supervision, Methodology, Funding acquisition, Conceptualization. \textbf{Oleg L. Polyansky} Conceptualization.
\textbf{Sergey N. Yurchenko} Supervision, Funding acquisition.

\section*{Declaration of competing interest}

The authors declare no conflict of interest.

\section*{Acknowledgments}
This work was supported by the UCL Center for Doctoral Training in Data Intensive Science, funded by the STFC training grant reference ST/P006736/1, as well as by the STFC Projects ST/Y001508/1 and UKRI/ST/B001183/1, and ERC Advanced Investigator Project 883830 (ExoMolHD).

\section*{Data Availability}
All data used for training are available in the cited papers, see Section~\ref{sec:data}. The code used in this work is publicly accessible in the GitHub project repository, \url{https://github.com/mbarnfield63/ML_Isotopologue_Extrapolation.git}. All updated energy levels for CO$_2$ have been used to update the ``Dozen'' line list states file available on the ExoMol website, \url{www.exomol.com}. The ML-corrected CO isotopologue levels, in cm$^{-1}$, are available as supplementary material.

\newpage
\appendix
\section{Reduced Mass Equations for CO$_2$}
\label{app:mu_eqns}
Effective reduced mass for symmetric stretch mode.
\begin{equation}
    \mu_1 = \left(\frac{mass_O^1\times mass_C}{mass_O^1 + mass_C}+\frac{mass_O^2\times mass_C}{mass_O^2 + mass_C}\right)\div 2
    \label{Eq:mu1}
\end{equation}\\

Reduced mass for bending mode (Oxygen-Oxygen pair).
\begin{equation}
    \mu_2 = \frac{mass_O^1 \times mass_O^2}{mass_O^1 + mass_O^2}
    \label{Eq:mu2}
\end{equation}\\

Effective reduced mass for asymmetric stretch mode.
\begin{equation}
    \mu_3 = \frac{(mass_O^1 + mass_O^2) \times mass_C}{mass_O^1 + mass_O^2 + mass_C}
    \label{Eq:mu3}
\end{equation}\\

Overall reduced mass of the triatomic system.
\begin{equation}
    \mu_{all} = \frac{mass_O^1 \times mass_C \times mass_O^2}{mass_O^1 + mass_C + mass_O^2}
    \label{Eq:mu_all}
\end{equation}\\

\section{Additional Features for Combined CO \& CO$_2$ Dataset. }
\begin{table}[H]
\label{app:features_CO}
\begin{tabular}{lrr}
\hline
\textbf{Feature Name} & \textbf{Description}                                                                & \textbf{Data Type} \\ \hline
$v$                   & Vibrational energy level                                                           & Float              \\
$\mu$                 & Reduced Mass                                                                        & Float              \\
$\mu$ ratio           & Ratio of isotopologue's $\mu$ to the parent isotopologue's $\mu$                      & Float              \\ \hline
\end{tabular}
\end{table}
All features not applicable to CO, e.g., AFGL/Herzberg/\TROVE quantum numbers, $\mu_{1-3}$, and secondary oxygen masses, were passed into the model as zeros.

\newpage
\bibliographystyle{elsarticle-num} 
\bibliography{References/journals_astro,
References/journals_iso,
References/journals_phys,
References/CO,
References/CO2,
References/exogen,
References/exoplanets,
References/jtj,
References/machine_learning,
References/ML,
References/methods,
References/PH3}

\end{document}